\documentclass[usenatbib]{emulateapj}%
\usepackage{amsmath}
\usepackage{graphicx}
\usepackage[unicode=true,pdfusetitle,bookmarks=true,bookmarksnumbered=false,bookmarksopen=true,breaklinks=true,pdfborder={0 0 0},backref=section,colorlinks=true]{hyperref}
\hypersetup{linkcolor=blue,citecolor=blue}

\shorttitle{SPATIAL VARIATIONS OF TURBULENT PROPERTIES IN H{\sc i} OF THE SMC}
\shortauthors{Nestingen-Palm et al.}

\begin{document}
\title{SPATIAL VARIATIONS OF TURBULENT PROPERTIES OF NEUTRAL HYDROGEN GAS IN THE SMALL MAGELLANIC CLOUD USING STRUCTURE FUNCTION ANALYSIS}
\author{David Nestingen-Palm\altaffilmark{1}, Sne\v{z}ana Stanimirovi\'{c}\altaffilmark{1}, Diego F. Gonz\'{a}lez-Casanova\altaffilmark{1}, Brian Babler\altaffilmark{1}, Katherine Jameson\altaffilmark{2}, Alberto Bolatto\altaffilmark{2}}
\email{sstanimi@astro.wisc.edu}
\altaffiltext{1}{Astronomy Department, University of Wisconsin-Madison, 475 North Charter Street, Madison, WI 53706-1582, USA}
\altaffiltext{2}{Astronomy Department and Laboratory for Millimeter-wave Astronomy, University of Maryland, College Park, MD 20742, USA}



\begin{abstract}

We investigate spatial variations of turbulent properties in the Small Magellanic Cloud (SMC) by using neutral hydrogen (H{\sc i}) observations. With the goal of testing the importance of stellar feedback on H{\sc i} turbulence, we define central and outer SMC regions based on the star formation rate (SFR) surface density, as well as the H{\sc i} integrated intensity. We use the structure function and the Velocity Channel Analysis (VCA) to calculate the power-law index ($\gamma$) for both underlying density and velocity fields in these regions. In all cases, our results show essentially no difference in $\gamma$ between the central and outer regions. This suggests that H{\sc i} turbulent properties are surprisingly homogeneous across the SMC when probed at a resolution of 30 pc. Contrary to recent suggestions from numerical simulations, we do not find a significant change in $\gamma$ due to stellar feedback as traced by the SFR surface density. This could be due to the stellar feedback being widespread over the whole of the SMC, but more likely due to a large-scale gravitational driving of turbulence. We show that the lack of difference between central and outer SMC regions can not be explained by the high optical depth HI. 

\end{abstract}


\section{Introduction}\label{introduction}


Interstellar turbulence is known to play an important role in the process of molecule formation, star formation, cosmic-ray propagation, and even large-scale galaxy dynamics \citep{elmegreen2004,mckee2007}.
Over the past two decades there have been many observational and numerical studies of turbulent properties in the interstellar medium (ISM). However, understanding what physical processes drive interstellar turbulence still remains an open question. There are many possible contenders, such as stellar feedback (\citealt{KimJ2001,deAv2005,Joun2006,Tamb2009,Shet2012,Fauc2013,Gris2017,Ostr2011,Stil2013}), gravitational instabilities \citep{Wada2002,Bour2010,Krum2016}, thermal instabilities \citep{Krit2002,Pion2004}, as well as Magneto-Rotational-Instabilities \citep{Balb1991,Sell1999,Pion2004}. 
Observational constraints of the preferential importance of these processes in different ISM environments are still lacking. In addition, observational evidence for the spatial variation of turbulent properties in the ISM is only starting to emerge.  

The goal of this paper is to search for spatial variations of turbulent properties and probe the importance of stellar feedback within the Small Magellanic Cloud (SMC). The SMC is particularly interesting for this study because of its proximity and existing high-resolution neutral hydrogen (HI) observations \citep{Stan1999} which have been used to study interstellar turbulence via several different statistical approaches. The SMC is a dwarf irregular galaxy located around 60 kpc away \citep{West1991}, and is part of a three galaxy system that also includes the Large Magellanic Cloud (LMC) and our own Milky Way (MW). In comparison to the MW, the SMC's heavy element abundance is around 5 times lower \citep{Kurt1998}, while its interstellar radiation field is on average at least 3 times stronger \citep{Lequ1989, Sand2010}.
The total H{\sc i} mass of the SMC is  $4.2 \times 10^{8}$ M$_{\odot}$ \citep{Stan1999}.

\citet{Stan1999} showed that a single power-law function was needed to fit the H{\sc i} spatial power spectrum (SPS) of the SMC for individual velocity channels, suggesting turbulent intensity fluctuations over a range of scales from 30 pc to 4 kpc. As no turnover was found in the SPS, even at the largest spatial scales, this was
suggestive of significant energy injections on scales larger than the size of the SMC.
\citet{Stan2001} showed that the SPS slope was slightly steeper when the whole H{\sc i} data cube was averaged into a single velocity slice, changing from $-2.8$ to $-3.3$. This change was in agreement with the theoretical expectations from \citet{lazarian2000} which showed that both density and velocity fluctuations affect H{\sc i} intensity. The velocity channel analysis (VCA), which involves SPS calculation for progressively thicker velocity channels, can be used to separate density and velocity contributions. As a result of this analysis, the H{\sc i} density in the SMC was found to have a power-law slope of $-3.3$, while the velocity field has a slope of $-3.4$, both being slightly more shallow than what is expected for Kolmogorov type turbulence. A slightly steeper velocity slope of $-3.85$ was estimated using the same HI data set but by applying the Velocity Coordinate Spectrum method \citep{Chep2015}. This method calculates a 1-D power spectrum along the radial velocity axis, while performing spatial averaging.


\citet{Mull2004} studied the H{\sc i} turbulence in the spatially adjacent Magellanic Bridge, the gaseous region between the SMC and the LMC formed by SMC-LMC-MW interactions. They divided the Bridge into four regions and applied the SPS and the VCA analysis on these individual regions. They found that the two southern regions have very similar turbulent properties like the SMC, while the north-east region (closer to the LMC) featured a systematically more shallow SPS slope. This was the first evidence that H{\sc i} turbulent properties can vary on kpc scales, and \citet{Mull2004} interpreted this result as being due to two large-scale gaseous arms being pulled from the SMC, due to gravitational interactions with the LMC, and having different age, distance, and physical properties.  

A different approach to studying spatial variations of turbulent properties was introduced by \citet{Burk2010}. By using a set of isothermal numerical simulations they established a correlation between the sonic Mach number (M$_{s}$) and the higher statistical moments (skewness and kurtosis) of the H{\sc i} column density. By inverting this relationship and calculating higher statistical moments of the H{\sc i} column density in the SMC, \citet{Burk2010} were able to map out the distribution of M$_{s}$ across the SMC. While about 90\% of the H{\sc i} was estimated to be subsonic or transonic, higher M$_{s}$ values (up to 4) were found in localized areas around the SMC bar. \citet{Burk2010} suggested that high M$_{s}$ regions could be caused by the shearing or tidal effects that the bar is experiencing relative to the surrounding, diffuse HI.  However, as discussed in \citet{Burk2010}, this result was based on using isothermal simulations and it remains to be seen whether these conclusions would persist when more realistic, multi-phase  numerical simulations are employed.


The \citet{Burk2010} method was later applied on a set of spiral galaxies from the H{\sc i} Nearby Galaxy Survey (THINGS) in \citet{Maie2016}. Generally, uniform statistical moments were found across galaxies without obvious correlation between moments and star-forming regions. However, this study has a resolution of about 700~pc and the moments could be tracing only large-scale turbulence. \citet{Stil2013} examined the H{\sc i} velocity dispersion  for a sample of dwarf galaxies and found that the line-width of the H{\sc i} super-profiles (calculated after rotation curves were removed) correlates with the star formation rate (SFR) and the H{\sc i} surface density, but the strongest correlation is found with the baryonic surface density. They examined several possible drivers of turbulence (thought to cause large line-widths) and found that stellar feedback is important but can not be the sole driver. They concluded that other physical mechanisms are important, but that it is also possible that the H{\sc i} line-widths are thermal in nature. Finally, \citet{Zhan2012} investigated the relationship between the HI SPS and star formation (calculated for entire galaxies) in a subsample of LITTLE THINGS dwarf irregular galaxies and found a lack of correlation between the SPS slope and the SFR surface density.


While many studies of the SPS exist for different regions in the MW, it is often difficult to directly compare the SPS slope due to different velocity resolution and the thickness of velocity slices \citep{Crov1983,Gree1993,Dick2000,Mivi2003,Khal2006} . We note a recent study by \citet{Ping2013} which addressed the issue of stellar feedback in particular. This study analyzed H{\sc i} observations of a non-starforming molecular cloud MBM16 in the MW using the SPS and VCA and found a steep density slope of $-3.7\pm$ 0.2. By comparing this result to other studies, \citet{Ping2013} suggested that the steep spectrum could be due to the lack of turbulent energy injection on small scales caused by stellar feedback. 
This study is important as it clearly observationally demonstrates that processes other than stellar feedback can result in a steep SPS slope.

In addition to observational studies, several recent numerical simulations of entire galaxies have searched for signatures of stellar feedback on the SPS slope.  Both \citet{Walk2014} and \citet{Gris2017} showed that stellar feedback significantly steepens the power spectrum slope by essentially destroying interstellar and giant molecular clouds, thereby shifting the power from small to larger scales. In particular, \citet{Gris2017} simulated H{\sc i} in an SMC-like galaxy and saw that the SPS slope (derived from the simulated H{\sc i} column density and therefore tracing the density field) changed from $\alpha=1.2$ to 1.7 (or from 2.0 to 3.0 at very small spatial scales) between no-feedback and feedback-included simulations. However, their SPS slope was always more shallow than what was found in observations, e.g. \citet{Stan1999}, and the lack of cosmological environment in the simulations could be a concern.
Finally, \citet{Krum2016} used the star formation rate and velocity dispersion averaged over whole galaxies to test the influence of stellar feedback as well as gravitational instabilities, and concluded that gravity-driven turbulence fits observations slightly better in galaxies with intense star formation. 
 



Clearly, observational signatures of the effect of stellar feedback on the SPS are currently lacking and are important to constrain numerical simulations. We therefore continue \citet{Stan1999}'s and \citet{Stan2001}'s work, but focus on sub-regions of the SMC where H{\sc i} is expected to be largely affected by recent star formation. We first reproduce \citet{Stan2001}'s SPS results for the entire SMC by using the structure function. Then, to analyze the effects of stellar feedback in the SMC, we delimit a central and outer region of the SMC based on the star formation rate, as well as the H{\sc i} integrated intensity. 

We use in this study the HI observations of the SMC from \citet{Stan1999}. This is still the highest resolution and the most sensitive HI data set for the SMC. As new HI observations of the SMC are anticipated in the near future with GASKAP \citep{Dick2013}, it is very timely to revisit the turbulent properties of the SMC and motivate the need for even higher resolution HI observations. The key novel aspect of our study is the search for spatial variations of turbulent properties in the SMC. Of all previous statistical studies of the SMC, only \citet{Burk2010} have addressed this issue. However, to calculate statistical moments they had to smooth the HI data to 30'. This effectively left them without adequate resolution to study turbulent properties close to star-forming regions. In addition, for the first time we are applying a structure function analysis on the HI data cube of the SMC. While the SPS and the structure function are directly related in theory, in practice only structure function can be computed for small SMC sub-regions as the SPS suffers from significant edge effects (we discuss this in Section 3).

This paper is organized in the following way.
In Section \ref{data}, we provide information about the data used in our structure function calculations. In Section \ref{methods} we present and discuss the structure function, its advantages, as well as other methods we use for calculations. Section \ref{results} discusses our results based on the SFR and H{\sc i} integrated intensity delimitation of the central and outer regions of the SMC. In Section \ref{opticaldepth} we discuss whether the correction for high optical depth can impact our results. Finally, we summarize our conclusions in Section \ref{conclusion}. 


\section{Observational Data}\label{data}

\subsection{H{\footnotesize I} data}

\begin{figure}[h]
\centering\includegraphics[width=\linewidth,clip=true]{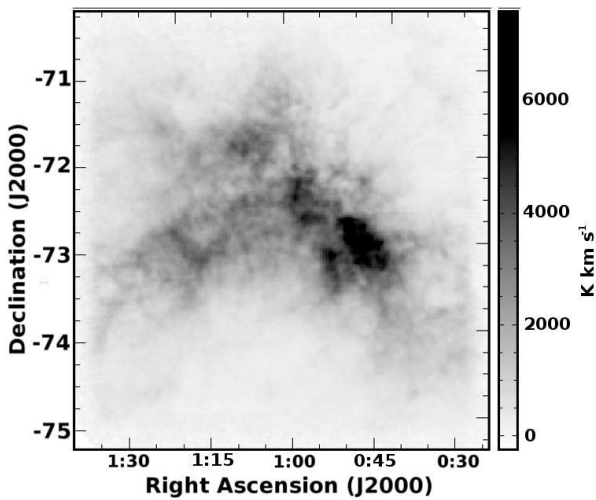}
\caption{The H{\sc i} integrated intensity image of the SMC \citep{Stan1999}. The H{\sc i} data cube is a combination of ATCA and Parkes observations and has an angular resolution of 98''.}
\label{SMCfullyInt}
\end{figure}

The neutral hydrogen (H{\sc i}) data used in this study are from \citet{Stan1999}. The H{\sc i} integrated intensity image can be seen in Figure \ref{SMCfullyInt}. The data are a combination of H{\sc i} observations of the SMC from the Parkes telescope and the Australian Telescope Compact Array (ATCA). 
The ATCA used five antennas in a 375 meter, East-West array configuration to observe 320 overlapping fields which covered 20 deg$^2$ and contained the SMC. The 64 meter Parkes Telescope was used to observe 1540 pointings centered on 01$^{h}$01$^{m}$, Dec. $-72^{\circ}$56' (J2000) and covered 4.5$^{\circ}$x4.5$^{\circ}$.  The combination of interferometer and single-dish observations was performed in the image domain and the combined `dirty' data cube was cleaned using MIRIAD's maximum entropy algorithm \citep{Saul1996}. The final H{\sc i} data cube has angular resolution of 98-arcsec and an H{\sc i} column density noise level of $4.2 \times 10^{18}$ cm$^{-2}$ per 1.65 km~s$^{-1}$ wide velocity channels, for more details please see \citep{Stan1999}. The data cube contains 78 velocity channels, each 1.65 km s$^{-1}$ wide.

\subsection{The star formation rate surface density}

\begin{figure}[h]
\centering\includegraphics[width=\linewidth,clip=true]{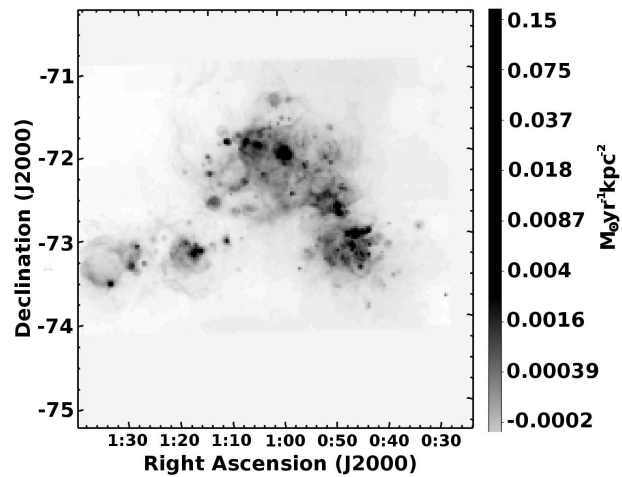}
\caption{The calculated star formation rate surface density of the SMC. The H$\alpha$ and 24-$\mu$m images were combined following Equation \ref{Jame2016Eqn} to create this image. The H$\alpha$ and 24-$\mu$m images were obtained from \citet{Bola2011} and \citet{Jame2016}.}
\label{SMCSFR}
\end{figure}

We use the H$\alpha$ and the 24~$\mu$m images from \citet{Bola2011} and \citet{Jame2016}, originally obtained from the Magellanic Cloud Emission Line Survey (MCELS; \citealt{Smit1999}) and the Spitzer Survey ``Surveying the Agents of Galaxy Evolution'' (SAGE; \citealt{Gord2011}) respectively. We follow \citet{Jame2016} and use H$\alpha$, which is corrected for extinction using the 24~$\mu$m image, to trace the star formation rate surface density of the SMC. The following prescription by \citet{Calz2007} is used to convert H$\alpha$ and 24~$\mu$m luminosities to the star formation rate \citep{Jame2016}:
\begin{multline}\label{Jame2016Eqn}
SFR(M_{\odot}~yr^{-1}) = 5.3 \times 10^{-42}[L(H\alpha) \\ 
+ (0.031 \pm 0.006)L(24\mu m)],
\end{multline}
where $L(H\alpha)$ and L(24~$\mu$m) are the luminosities of H$\alpha$ and 24~$\mu$m images, respectively, and are measured in erg~s$^{-1}$. This SFR image is in essence the same image as shown in Figure 1 of \citet{Bola2011}. The resulting rms noise value is 4$\times$10$^{-4}$ M$_{\odot}$~yr$^{-1}$~kpc$^{-2}$ \citep{Jame2016}. The correction by the 24~$\mu$m image amounts to a $\sim$10$\%$ contribution to the total star formation rate image \citep{Jame2016}. 
Because of this, we only use the SFR image in our analysis as results obtained using the H$\alpha$ image are very similar. The SFR image is shown in Figure \ref{SMCSFR}.

\section{Methods}\label{methods}
We use the structure function as our method for analyzing the H{\sc i} distribution and turbulent properties in the SMC. While several previous studies have used the spatial power spectrum, apodizing approaches are often needed due to strong edge effects when Fourier transforming the images. Common apodizing approaches involve placing the image in a larger array padded with zeros and/or the image edges being slightly smoothed to ensure a gradual decrease in intensity \citep{Stan1999,Burk2010}. While these are widely used methods in signal processing, they involve decisions about the smoothing kernel size and/or padding size. 

The structure function, however, stays within the image domain, which makes analysis and calculations more straight forward and also slightly easier to comprehend. As Fourier transforms are not applied, it is straight forward to analyze smaller subsections of an image without worrying about apodizing kernels. Finally, if the spatial power spectrum of a distribution is a power law function with a spectral index $\alpha$ (such that SPS $\propto r^{-\alpha}$), then the structure function will also be a power law with an index $\gamma$, such that SF $\propto r^{-\gamma}$. If $2<\alpha<4$, then the two indices are related by the equation \citep{Simo1984}:
\begin{equation} \label{StructurePowerRelation}
\gamma =\alpha-2.
\end{equation}
Therefore, when working with a structure function we fit a power-law function whose index can be directly related to the spectral index of the spatial power spectrum. We note that the structure function has been used extensively for the study of turbulence in the ISM (please see references in \citet{Have2004}).  

\begin{table}
\begin{center}
\caption{}
\label{CubeSliceThicknesses}
\begin{tabular} { |c|c| }
\hline
\# of Channels Averaged &  $\Delta v$ (km~s$^{-1}$) \\
\hline\hline
1 & 1.65 \\
\hline
2 & 3.30 \\
\hline
3 & 4.95 \\
\hline
6 & 9.89 \\
\hline
13 & 21.4 \\
\hline
26 & 42.9 \\
\hline
39 & 64.3 \\
\hline
78 & 129 \\
\hline
\end{tabular}
\bigskip\\
Summary of 8 PPV data cubes used in this study. The original data cube corresponds to the top row of the table and has a velocity slice thickness of 1.65 km s$^{-1}$. By averaging the number of velocity channels shown in the first column, we derive the velocity channel thickness $\Delta v$ shown in the second column.
\end{center}
\end{table}

We calculate structure functions using the following equation:
\begin{equation} \label{StructureFuncEqn}
SF_{I}(r)=\left \langle[I(r')-I(r'+r)]^{2} \right \rangle_r' \vert_{\Delta v}
\end{equation}
Here, $I$ represents brightness (this is brightness temperature in the case of the SMC H{\sc i} data cube), $r'$ represents an arbitrary pixel in the image, $r$ is the distance lag used for calculations. For a given lag $r$, a squared difference in image brightness is calculated and averaged over all pixels $r'$. The bar $\Delta v$ shows that the structure function is calculated for a specific thickness of velocity channel. 

Working with a data cube, we can vary $\Delta v$ and we use this to disentangle influence of velocity vs density fluctuations on the brightness structure function. Therefore, we calculate the structure function for the H{\sc i} integrated intensity image and for individual H{\sc i} cube velocity channels. We also progressively average velocity channels together to lower velocity resolution, perform the structure function calculation on these new individual velocity channels, and average the resultant slopes. As mentioned in the introduction, this method is called the velocity channel analysis (VCA) and was originally utilized by \citet{lazarian2000}. Table \ref{CubeSliceThicknesses} shows the resultant velocity resolution for the differing velocity thicknesses, starting with $\Delta v=1.65$ km s$^{-1}$ when we use individual velocity channels, 
all the way to $\Delta v=129$ km s$^{-1}$ when we average all 78 velocity channels.

For calculating structure functions we first rebin the H{\sc i} data cube so that each pixel has angular size equal to the telescope beam of 98$''$ to ensure that pixels are independent. When plotting the structure functions we convert the angular separation into linear size by assuming a distance of 60 kpc \citep{West1991}. We sample the range of scales from about 40 pc to the maximum length allowed by an image under consideration. For example, when using the entire SMC image the maximum scale is 2 kpc, which corresponds to one half of the image's shortest side. When working with smaller image sub-sections, the maximum size is always set to about one half of the largest scale that is perpendicular to the largest possible scale of the image. As an example, if the smaller image was an ellipse, the maximum scale would be the semi-minor axis, as this is half of the largest scale (minor axis) that is perpendicular to the largest possible scale (major axis). When working close to the image edge, if $I(r'+r)=0$ we exclude $I(r')-I(r'+r)$ from our calculation. This effectively lowers the number of difference-squared values used in equation (3), however does not affect the structure function as the number of pairs is generally very large (hundreds of thousands).

We note that structure function calculations naturally lead to \textit{double dipping} of measurements. In other terms, when getting the difference-squared value for data-pixel $A$ when compared to $B$, the same value is counted again when data-pixel $B$ is compared to $A$. This can lead to an artificial increase in statistics and longer computation times. We avoid this effect in our calculations by allowing the difference-squared values to be calculated only from pixels with an angular separation ranging from 0 to $<180$ degrees relative to the
reference pixel.
Finally, it was pointed out by Emmanoulopoulos et al. (2010) that derivation of the structure function slope from a 1-D data set can be biased by gaps in the data. This is not a concern in our study as we mainly use well-sampled, connected regions for analysis. To test whether image masking affects the structure function slope, we performed simple simulations of a Gaussian random field with a specified structure function slope 
and confirmed that masking is not an issue when the range of scales for calculating the slope is carefully estimated as discussed above (we specify the exact range of scales used for our calculations in Section 4).

\begin{figure}[h]
\centering\includegraphics[width=\linewidth,clip=true]{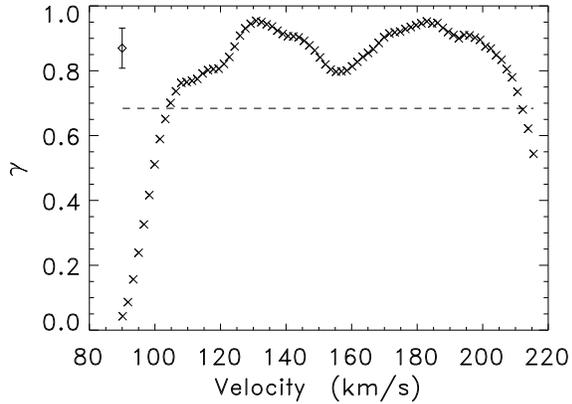}
\caption{The power-law index ($\gamma$) of the HI structure function for each individual velocity slice (at $\Delta v=1.65$ km~s$^{-1}$) of the SMC data cube. The dashed line represents the sigma clipping cut-off used to estimate the average power-law index $\langle \gamma \rangle$ shown in the top left corner and calculated as the mean and the standard deviation of all points above the dashed line.} 
\label{1regStddevSlope78slice}
\end{figure}

We create 8 position-position-velocity (PPV) data cubes from our original H{\sc i} data cube, each with a different velocity channel thickness ($\Delta v$) as shown in Table \ref{CubeSliceThicknesses}. 
This number of PPV cubes was derived to ensure averaging by an integer number of velocity channels only\footnote{ There are 8 integers that divide into the 72 slices evenly, these being 1, 2, 3, 6, 13, 26, 39, and 72. For the last cube we have averaged all 78 channels instead of just 72. By averaging only an integer number of channels we have ensured that all channels in each one of the 8 PPV cubes are statistically independent.  }.
For each cube we analyze each velocity slice separately and apply the structure function, therefore calculating a power-law slope ($\gamma$) for each velocity slice by performing a simple polynomial fit.  As an example, Figure \ref{1regStddevSlope78slice} shows the $\gamma$ values for each velocity slice for the PPV cube with $\Delta v$ = 1.65 km~s$^{-1}$. We then estimate the average slope, $\langle \gamma \rangle$, for each PPV cube. As shown in Figure \ref{1regStddevSlope78slice}, 
$\gamma$ values for most velocity channels have a reasonably constant value, while few velocity channels at the start and the end of the PPV cube are largely dominated by noise and have lower $\gamma$ values. In other words, if we plot a histogram of $\gamma$ values, points corresponding to the noisy channels represent a tail of a strongly-peaked distribution. To exclude the tail from our mean calculation we apply the sigma clipping procedure.  This is an iterative process which starts by calculating the median value and the standard deviation of all $\gamma$ values, and then removes outliers beyond 2.5-$\sigma$ boundaries. The process is repeated several times (converges very fast in our case) until no more outliers are present. From the set of remaining $\gamma$ values we estimate the mean and standard deviation which gives us $\langle \gamma \rangle$ and the error for $\langle \gamma \rangle$ respectively. The dashed line in Figure \ref{1regStddevSlope78slice} shows all points above the line which was used to calculate $\langle \gamma \rangle$ as the result of the sigma clipping procedure.


\section{Results: Spatial Variations of Turbulent Properties}\label{results}

\subsection{Reproducing power spectrum analysis}

\begin{figure}[h]
\centering\includegraphics[width=\linewidth,clip=true]{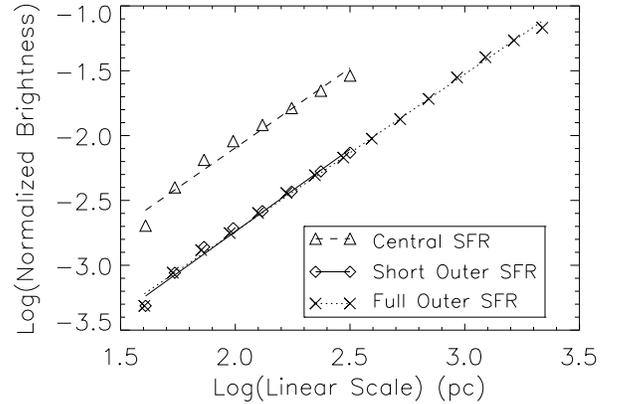}
\caption{The structure function of the entire integrated H{\sc i} image (shown with crosses) and its
power-law fit, resulting in $\gamma=1.25\pm0.04$. The fit was performed over the entire range of scales from 40 pc to 2 kpc. The structure functions of the H{\sc i} integrated intensity image calculated for the central and outer SMC regions, as defined in Section 4.2.1, are shown with triangles and diamonds, respectively.  The central region has the slope $\gamma=1.23\pm0.07$, while the slope of the outer region is $\gamma=1.27\pm0.04$. The structure function of the central and outer SMC regions was fitted over exactly the same range of spatial scales, 40 pc to 370 pc, for direct comparison. The y-axis is unitless because we normalize the data over the maximum brightness temperature value before preforming the structure function calculation.}
\label{TwoStrucFunc}
\end{figure} 


We start with the entire integrated H{\sc i} image (corresponding to $\Delta v=129$ km s$^{-1}$) and derive the structure function slope to check that our results agree with previous studies. We estimate the structure function slope $\gamma$=1.25$\pm$0.04\footnote{The error was calculated by using a Monte Carlo simulation to propagate random noise in each data pixel, as well as a bootstrap analysis to estimate the power-law slope of the structure function.}, which corresponds to the power spectrum slope of $\alpha$=$-3.25$$\pm$0.04. This is in excellent agreement with previous studies \citep{Stan2000,Stan2001,Burk2010}. Figure \ref{TwoStrucFunc} shows the structure function for the entire SMC image, as well as the central and outer SMC regions we discuss later.

\begin{figure}[h]
\centering\includegraphics[width=\linewidth,clip=true]{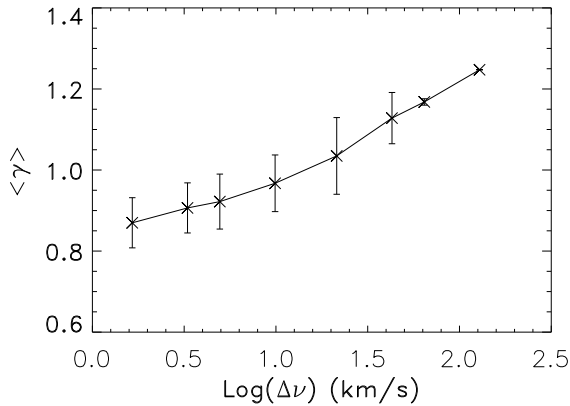}
\caption{The average structure function slope ($\langle \gamma \rangle$) as a function of the thickness of velocity slices derived for the entire SMC area. The average values are calculated by using a sigma clipping method.}\label{SMCGammaGraph}
\end{figure}

We next calculate the structure function slope for each velocity channel with $\Delta v$ = 1.65 km~s$^{-1}$, our thinnest velocity slices. These slopes are shown in Figure \ref{1regStddevSlope78slice} and are again in agreement with \citet{Stan1999}. To estimate $<$$\gamma$$>$ at this velocity resolution, we preform a sigma clipping of all data points in Figure \ref{1regStddevSlope78slice} as discussed in Section~\ref{methods}.
Figure \ref{SMCGammaGraph} shows $<$$\gamma$$>$ from the above calculations, as well as $<$$\gamma$$>$ obtained by averaging the H{\sc i} data cube to intermediate velocity resolutions (e.g. progressively averaging every 2, 3, 6, etc velocity channels, please see Table \ref{CubeSliceThicknesses} for more details). Taking into consideration Equation \ref{StructurePowerRelation}, the results shown in this figure match quite well to the results shown in \citet{Stan2001}'s Figure 1. This shows that our structure function calculations are working well and we are able to reproduce previous results. As explained in \citet{lazarian2000}, the gradual change in the power-spectrum or structure-function slope shows that H{\sc i} intensity images have contribution from both density and velocity fluctuations, with density being the dominant contribution at the highest $\Delta v$ end and velocity fluctuations dominating at the lowest $\Delta v$ end. Numerical simulations of stellar winds by \citet{Offn2015} have shown that winds influence significantly the velocity power spectrum, while the density power spectrum is less affected. In our investigation of the turbulent properties of different SMC sub-regions we therefore apply the VCA technique to search for changes in both density and velocity fluctuations.

\subsection{Spatial variations of turbulent properties}

\subsubsection{Central star-forming SMC}

The two dominant turbulent drivers discussed in the literature are gravity (via gravitational instability coupled with galactic rotation) and stellar feedback (via stellar winds, ionizing radiation and supernova explosions). In the first approximation, these two different turbulent drivers operate or dominate at different scales.  
For example, as star formation is concentrated in central regions of galaxies we would expect the most significant influence of stellar feedback to be in shaping velocity/density structure in H{\sc i} regions close to star-formation sites. 
Numerical simulations by \citep{Gris2017} suggest that stellar feedback affects smaller spatial scales, but that the shape of the SPS on spatial scales $>1$ kpc is insensitive to stellar feedback. On the other hand, gravitational instabilities due to galactic rotation will likely be observable over the entire SMC area.  Even larger-scale gravitational effects and instabilities associated with galaxy interactions (e.g. SMC-LMC-MW system) may show observational signatures on scales larger than the size of the SMC.

\begin{figure*}[h]
\centering\includegraphics[width=\linewidth,clip=true]{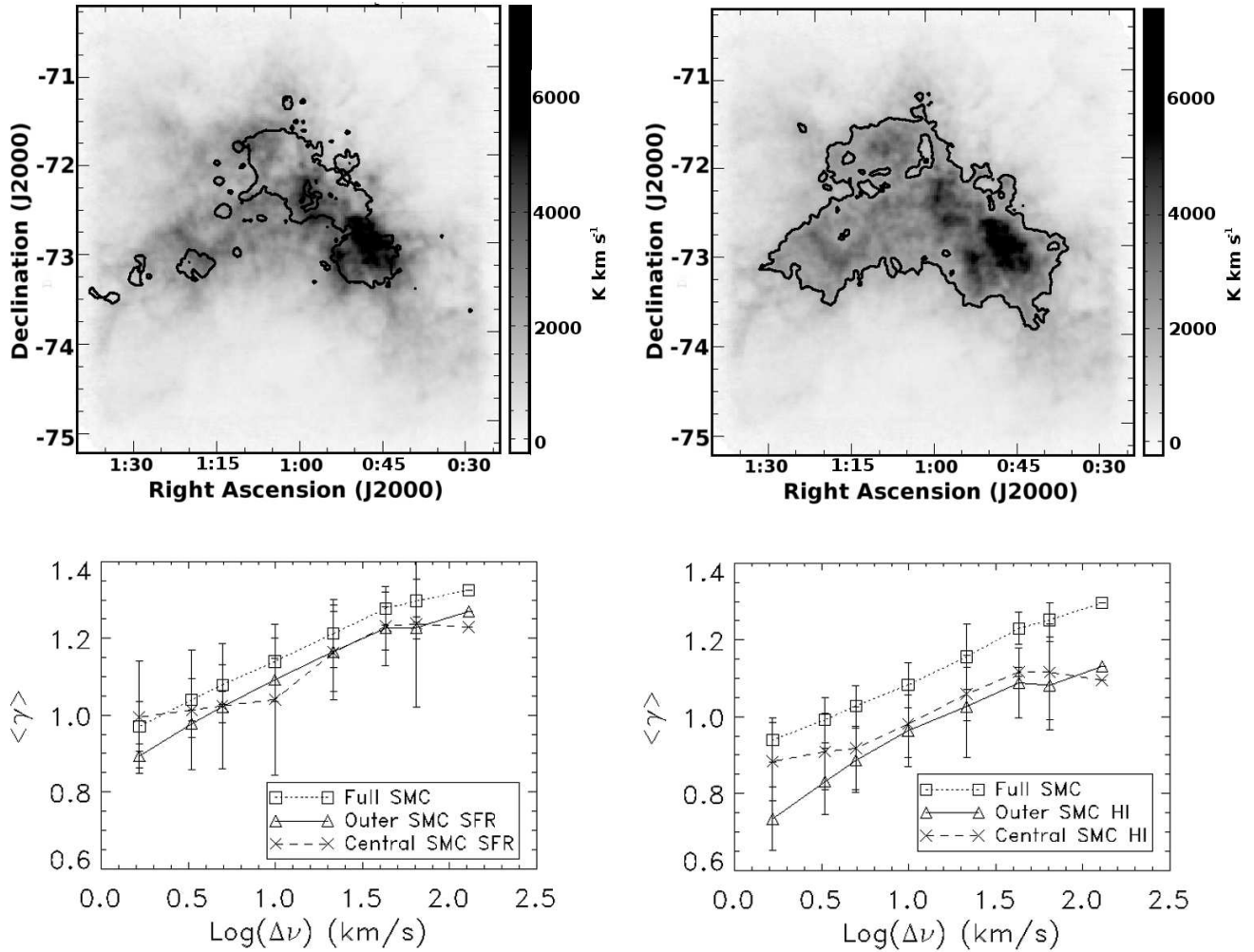}
\caption{\textbf{Top Left:} The H{\sc i} integrated intensity image of the SMC overlaid with the SFR contour of 0.004 M$_\odot$~yr$^{-1}$~kpc$^{-2}$ used to define our central and outer SMC regions based on the star formation rate.  \textbf{Top Right:} The H{\sc i} integrated intensity image overlaid with an H{\sc i} contour level of 2000 K km~s$^{-1}$ to distinguish the highest H{\sc i} intensity region of the SMC. 
\textbf{Bottom:} (Left) The average structure function slope as a function of the thickness of velocity slices for the outer and central SMC regions defined by the star formation rate contour level of 0.004 M$_\odot$ yr$^{-1}$ kpc$^{-2}$. The structure function slope was calculated over the range of 40 pc to 370 pc. (Right) The average structure function slope for the central and outer SMC as defined with the H{\sc i} integrated intensity contour of 2000 K km~s$^{-1}$. The structure function slope was calculated over the range of 40 pc to 660 pc.}\label{ContoursandGammas}
\end{figure*}

To search for the influence of stellar feedback on the HI turbulence, we divided the SMC into a central and outer region by using the SFR image and by assuming that stellar feedback correlates with regions of the highest SFR. 
In Figure \ref{ContoursandGammas} (top left) we over-plot the contour level corresponding to the SFR of 0.004 M$_{\odot}$~yr$^{-1}$~kpc$^{-2}$ (this corresponds to 10 times the rms background noise of the image, \citet{Jame2016}), on the H{\sc i} integrated intensity image of the SMC. We use this contour level to create a central and outer SMC region and calculate the H{\sc i} structure function for both regions as we did for the entire SMC data cube. 
In the same manner, we progressively increase the thickness of velocity slices and repeat structure function calculations (all error calculations are done in exactly the same way as for the full SMC data cube). The results are shown in Figure \ref{ContoursandGammas} (bottom left) \footnote{We note that when calculating the structure function for the central region, the longest lag used for the structure function was calculated by approximating roughly the contour as an ellipse, and taking the semi-minor axis. For the outer region, we use the same maximum lag as is used on the central region for a direct comparison of results.}. The same analysis was performed for a SFR threshold of 0.006 M$_{\odot}$~yr$^{-1}$~kpc$^{-2}$ and results were unchanged.

Figure \ref{ContoursandGammas} (bottom left) shows that there is essentially no difference between the central and the outer SMC based on the SFR threshold over the whole range of $\Delta v$ values probed. This suggests that H{\sc i} in the central SMC, where most of the recent star formation is located (within last 3-10 Myrs, \citet{Kenn2012}), has very similar turbulent properties as the H{\sc i} in the outer SMC region.  This is a surprising result considering that several recent numerical simulations suggested significant change in the SPS column density slope due to stellar feedback. For example, both \citet{Walk2014} and \citet{Gris2017} showed that stellar feedback steepens the power spectrum slope by essentially destroying interstellar clouds, thereby shifting the power from small to larger scales. 
In their simulation of H{\sc i} in a SMC-like galaxy, \citet{Gris2017} found that the HI column density SPS slope changed from $\alpha=1.2$ to 1.7 (or from 2.0 to 3.0 at very small spatial scales) between no-feedback and feedback-included simulations. The \citet{Gris2017} simulations included effects of stellar winds and supernova explosions, and their stellar feedback simulations have a 10 times higher star formation efficiency than the no-feedback ones. 

We do not find observational evidence for such significant slope difference when comparing H{\sc i} within regions of the most intense and recent stellar activity and 
the H{\sc i} contained in the rest of the SMC.
In addition, our inner SMC region contains about 1 kpc in length of the main star-forming body of the SMC. This is well within the range where \citet{Gris2017} found strong signatures of stellar feedback in their simulated galaxies. Our results (this and next sub-section) suggest relatively uniform turbulent properties throughout the SMC.
In addition, no significant difference is found for either thick velocity slices (probing density fluctuations) or thin velocity slices (probing velocity fluctuations). The error bars of $\langle \gamma \rangle$ for thin velocity slices are larger than those of thick velocity slices. The reason for this are small variations in the slope derived for different velocity channels as shown in Figure 3.


The structure function of the central and outer SMC regions suggests a corresponding SPS slope of $-3.23\pm0.07$ and $-3.27\pm0.04$, respectively, at $\Delta v=129$ km~s$^{-1}$. 
Within the \citet{Gris2017} numerical framework, 
such steep density slope is seen only if stellar feedback is included. It is therefore tempting to conclude that the influence of stellar feedback is wide spread across the SMC, reaching even to far galactic outskirts. However, we note that a steep SPS density slope ($-3.7\pm0.2$) was also measured in a starless interstellar cloud MBM16 by \citet{Ping2013}, suggesting that other physical processes (likely shear due to Galactic rotation in the case of MBM16) can also produce a steep SPS. Therefore, the observed turbulent spectrum in the SMC could be caused all together by other drivers, such as large-scale shearing due to galaxy-galaxy interactions or SMC's rotation relative to the surrounding low-density HI in the Magellanic Bridge. 
This is a very likely explanation especially when considering that the Magellanic Bridge, which only has traces of internal star formation, shows a SPS density slope similar to that of the SMC. As \citet{Mull2004} showed, two southern Bridge regions adjacent to the SMC and extending up to an additional 4 kpc from the SMC's tail, have a very similar SPS density slope. As it is unlikely for SMC's stellar feedback to reach so far out from star-formation sites, this again supports the idea that turbulence in the SMC is driven on very large scales. This result agrees with the Velocity Coordinate Spectrum analysis by \citet{Chep2015} who estimated the turbulent injection scale of 2.3 kpc and suggested that the largest HI shells in the SMC or tidal interactions could be driving the turbulent cascade.

\subsubsection{Central high H{\sc i} integrated intensity SMC}

If instead of SFR we use the H{\sc i} integrated intensity to define the central vs outer SMC regions, we retrieve essentially the same result. This is shown in Figure~\ref{ContoursandGammas} (bottom right). As shown in Figure \ref{ContoursandGammas} (top right), we define a dividing contour to enclose the highest integrated intensity as the central region.  This contour was defined approximately as an average of the full width at half power of the integrated intensities obtained along RA and Dec axes at the position of the kinematic center of the SMC (defined in \citealt{Stan2004}). In terms of H{\sc i} mass, the central region contains $\sim$1.92$\times$10$^{8}$ M$_{\odot}$ relative to $\sim$1.94$\times$10$^{8}$ M$_{\odot}$ contained in the outer SMC. 

We again calculate the structure function for different thickness of velocity slices, the result is shown in Figure \ref{ContoursandGammas} (bottom right). We find no difference in $<$$\gamma$$>$ between the inner and central SMC regions when this definition is used. Similarly to the SFR, this suggests that the SMC is very uniform regarding its HI turbulence properties. Looking at the outer-H{\sc i} region in this case, which focuses exclusively on more outer areas than in the previous section (at least 1 kpc away from the SMC center), we still find a steep SPS slope with the density spectral index of $-3.13\pm0.05$.  This result is in agreement with the power-spectrum analysis \citep{Stan1999,Stan2001,Gold2000} where no turnover on the largest sampled scales was noticed. This was interpreted as being due to the energy injection on scales larger than the size of the SMC. We note that we have used a range of H{\sc i} integrated intensity cut-off values and our results are consistent. 

\citet{Burk2010} estimated the spatial distribution of the sonic Mach number (M$_{s}$) across the SMC using the same H{\sc i} data set and by applying the relationship between skewness and kurtosis of the H{\sc i} column density and the M$_{s}$ from isothermal numerical simulations. They noticed localized areas with enhanced M$_{s}$ just off the SMC bar region and suggested that large-scale shearing, caused by gravitational interactions between the SMC, LMC and the MW, could be responsible for spatial variations of turbulent properties. Our H{\sc i} contour used to distinguish between the central and outer SMC includes the SMC bar and the region just off the bar with enhanced M$_{s}$. However we do not find significant difference in structure function slopes between the central and outer SMC. This could suggest that the enhanced M$_{s}$ regions are highly localized and therefore do not show up in structure function calculations which average many pixel pairs at a given lag.

\section{High optical depth of H{\sc i}}\label{opticaldepth}

\begin{figure}[h]
\centering\includegraphics[width=\linewidth,clip=true]{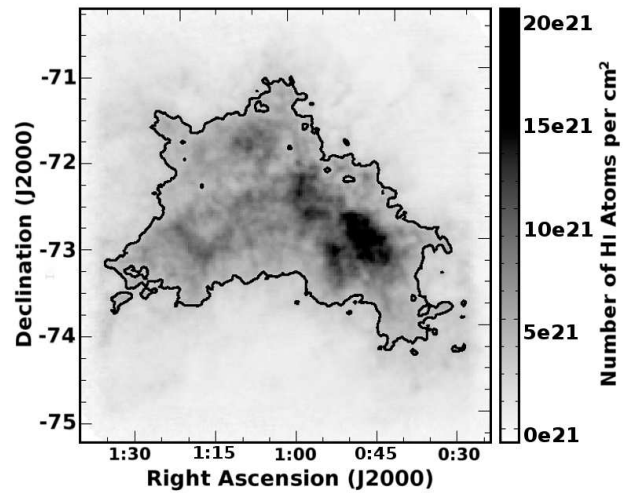}
\caption{The H{\sc i} column density image corrected for high optical depth using Equation \ref{OpticalDepthEqn}. Overlaid on top of the image is a contour of $>$10$^{21.4}$ cm$^{-2}$, all pixels inside this contour were corrected for the high optical depth.}\label{HighOpticalDepthFix}
\end{figure}

The H{\sc i} data cube we use in our analysis is not corrected for high optical depth. Regions with high H{\sc i} optical depth are likely to be found close to star-forming areas of the SMC and this could affect the power spectrum and structure function results. \citet{Stan1999} provided a correction factor that should be applied on the H{\sc i} column density image of the SMC to correct for high optical depth for all pixels with H{\sc i} column density greater than 10$^{21.4}$ cm$^{-2}$. This is shown in Equation \ref{OpticalDepthEqn} where $f_{c}$ is a factor that the H{\sc i} column density needs to be multiplied by. The region with column density larger than $10^{21.4}$ cm$^{-2}$ is quite extensive so the correction could be significant, see Figure \ref{HighOpticalDepthFix}. 

\begin{equation} \label{OpticalDepthEqn}
f_{c} = \left\{
\begin{array}{ll}
      1 + 0.667(logN_{H_{I}} - 21.4) & logN_{H_{I}} > 21.4 \\
      1 & logN_{H_{I}}\leq 21.4 \\
\end{array} 
\right.
\end{equation}

As the correction can be applied only on the HI column density, we test the effect of high optical depth only for the case of $\Delta v=129$ km~s$^{-1}$ and the slope of density fluctuations. We apply the above correction and then calculate the structure function slope for the entire SMC area as well as the central and outer regions, see Table \ref{HICorrectedGammaTable} for results. All of the resultant density slopes are very close to, if not the same as, their previous values. This shows that the lack of high optical depth correction can not explain our results in Section 4 for the slope of density fluctuations (obtained at $\Delta v=129$ km s$^{-1}$). 
However, it remains to be tested with future HI observations whether the 
the correction for high optical depth, when applied on the individual velocity channels, can affect the 
slope of velocity fluctuations (in the case of $\Delta v=1.65$ km s$^{-1}$).
We note that \citet{Laza2004} showed that the SPS slope of intensity fluctuations derived for an absorbing medium saturates at $-3$. Figure 6, where we do not find slope saturation suggests that the HI in the SMC is largely optically thin.


\begin{table}[h]
\begin{center}
\caption{}
\label{HICorrectedGammaTable}
\begin{tabular} { |c|c|c| }
\hline
Region & Original $\gamma$ & Corrected $\gamma$ \\
\hline\hline
Entire SMC & 1.25 & 1.21 \\ 
\hline
Central SFR & 1.23 & 1.23 \\
\hline
Outer SFR & 1.27 & 1.26 \\
\hline
Central H{\sc i} & 1.10 & 1.10 \\
\hline
Outer H{\sc i} & 1.13 & 1.09 \\
\hline
\end{tabular}
\bigskip\\
The structure function slope before and after the high optical depth correction is applied to the H{\sc i} column density. These SPS slopes correspond to the 3D density field. An average error of $\pm$0.05 is calculated for these power indicies by a combination of Monte Carlo simulations and boot-strap analysis.
\end{center}
\end{table}


\section{Discussion and Conclusions}\label{conclusion} 

We have searched for spatial variations of H{\sc i} turbulent properties in the SMC, with the goal of probing the importance of stellar feedback, by employing the structure function. When used on the entire H{\sc i} data cube, our structure function analysis confirms previous SPS findings for individual velocity channels and also progressive averaging of velocity channels \citep{Stan1999,Stan2001}.  We then selected specific regions in the SMC to test the importance of stellar feedback for turbulent properties. We divided the SMC into the central and outer regions by using the SFR image, as well as the H{\sc i} integrated intensity, to select HI regions dominated by the recent stellar feedback and H{\sc i} regions of the highest integrated intensity. To measure the turbulent spectral index in those four regions we employed the VCA and the structure function analyses. 

We have found that there is virtually no difference in HI turbulent properties between regions of high SFR vs low SFR, as well as the high vs low H{\sc i} integrated intensity. 
The H{\sc i} in the SMC appears to have relatively uniform turbulent properties.  Several numerical simulations on galaxy-large scales suggest a significant change in the 
SPS density slope due to stellar feedback e.g. \citet{Walk2014,Gris2017}, 
suggesting that at least a slight change in the SPS slope should be found in the vicinity of the most recent star formation activity.
However, our observations do not find a change in the SPS slope (density or velocity) when we use the SFR as a tracer of the most intense and recent stellar feedback. Considering that the SPS density slope of the SMC is steep, consistent with what is hinted by feedback-dominated numerical simulations, one possible interpretation is that the stellar feedback influence is wide-spread across the SMC. However, a steep SPS can be produced by non-stellar processes as well (e.g. \citet{Ping2013,Bour2010}). In addition, the adjacent Magellanic Bridge has a SPS density slope similar to what we find in the SMC, suggesting a large-scale turbulent driving. This conclusion agrees with the recent modeling of the LMC by \citet{Bour2010}, who found that gravitational instabilities alone can reproduce the observed SPS. Our observed lack of change in the structure function slope in the central star-forming SMC agrees with recent results by \citet{Zhan2012}, who for a sample of irregular dwarf galaxies found that the SPS slope does not correlate with the SFR surface density. They concluded that either non-stellar sources are more important in driving turbulence, or that turbulent properties have nothing to do with their initial driving sources.  

One alternative possibility, however,  is that the SFR is not a good tracer of the recent stellar feedback and that we have been simply searching for the influence of stellar feedback in wrong places. For example, \citet{Stil2013} suggested that due to a commonly bursty nature of star formation in dwarf galaxies, the SFR may not be the best tracer of the HI turbulence affected by the recent star formation. This argument probably does not apply in the case of the SMC as \citet{Harr2004} showed a relatively uniform star formation history over the last 60 Myrs. In addition, the spatial distribution of the SFR and supernova remnants are relatively similar and predominately along the SMC bar (e.g. \citet{Fili2005}).

While we have compared our results with numerical simulations of galaxies which self-consistently include stellar feedback (supernovae and stellar wind), we note that studies on smaller spatial scales have also investigated effects of turbulent driving on multiple scales. For example, \citet{Yoo2014} performed analytic calculations, as well as hydrodynamic and MHD simulations with and without magnetic field by driving turbulence at two spatial scales. They showed that the energy spectrum is very sensitive to large-scale driving, even a small amount of energy injection on large scales can change the turbulent spectrum. This change should be observed in the kinetic energy spectrum, but also density and column density spectra. In addition, the small-scale driving does not affect turbulent properties unless energy injection rates on small and large scales are comparable. 
While it is currently not possible to connect these results with those of large-scale galaxy simulations (e.g. \citet{Walk2014,Gris2017}) which appear highly sensitive to stellar feedback, future numerical studies that bridge small- and large-scale simulations are essential to understand the role of turbulent driving at multiple scales. At the same time, further observational studies are essential to test and constrain numerical simulations.

We have also tested whether the high optical depth could be responsible for the lack of difference between the central and outer SMC regions when considering the slope of density fluctuations. From applying the correction for high optical depth on the H{\sc i} column density we see that this is not the case. To apply the high optical depth correction on the full H{\sc i} data cube we would require H{\sc i} absorption spectra in many directions, this will likely be possible in the near future with GASKAP \citep{Dick2013}. Finally, another possibility is that regions of enhanced turbulent properties are highly localized and therefore do not stand out in the structure function calculations. This can also  be tested in the future with GASKAP and higher resolution HI observations. In addition, with the future higher velocity resolution observations we can likely probe better the thin-slice portion of the structure functions in Figure 6 and decrease the size of error bars.

\acknowledgments
We thank an anonymous referee for constructive comments and suggestions.
We thank Jay Gallagher and Claire Murray for stimulating discussions. This work was supported by the NSF Early Career Development (CAREER) Award AST-1056780. Partial support for DFGC was provided by CONACyT (M\'{e}xico).

\bibliographystyle{apj.bst}
\bibliography{Biblio}

\end{document}